\documentclass[fleqn,twoside]{article}
\usepackage{espcrc2}

\usepackage{graphicx}

\def\lsim{\mathrel{\rlap{\lower4pt\hbox{\hskip1pt$\sim$}}
    \raise1pt\hbox{$<$}}}         
\def\gsim{\mathrel{\rlap{\lower4pt\hbox{\hskip1pt$\sim$}}
    \raise1pt\hbox{$>$}}}         

\newcommand{\AmS}{{\protect\the\textfont2
  A\kern-.1667em\lower.5ex\hbox{M}\kern-.125emS}}

\title{CP Violation: The CKM Matrix and New Physics}
\author{Yosef Nir\address[WIS]{Department of Particle Physics,
        Weizmann Institute of Science, Rehovot 76100, Israel}%
        \thanks{supported by the Israel Science Foundation founded by
          the Israel Academy of Sciences and Humanities.}%
        \thanks{plenary talk given at the 31st international
        conference on high energy physics (Amsterdam, $24-31$
        July, 2002).}
        }
       
\begin{document}

\begin{abstract}
Recent measurements of CP violating asymmetries have led to a
significant progress in our understanding of CP violation. The
implications of the experimental results for the Kobayashi-Maskawa
mechanism and for new physics are explained.
\vspace{1pc}
\end{abstract}

\maketitle

\section{Introduction}
The study of CP violation is, at last, experiment driven. 
Experiments have measured to date three independent CP violating
parameters:

$\bullet$ Indirect CP violation in $K\to\pi\pi$ \cite{Christenson:fg} and
  in $K\to\pi\ell\nu$ decays is given by
\begin{equation}\label{epskwa}
|\varepsilon|=(2.28\pm0.02)\times10^{-3}.
\end{equation}

$\bullet$ Direct CP violation in $K\to\pi\pi$ decays is given by
\begin{equation}\label{epspwa}
{\cal R}e(\varepsilon^\prime/\varepsilon)=(1.66\pm0.16)\times10^{-3}.
\end{equation}
(The world average given in eq. (\ref{epspwa}) includes the new result
from NA48 \cite{unalichep}, ${\cal
  R}e(\varepsilon^\prime/\varepsilon)=(1.47\pm0.22)\times10^{-3}$, and
previous results from NA31, E731 and KTeV.)

$\bullet$ The CP asymmetry in $B\to\psi K_S$ decay (and other, related,
  modes) has been measured:
\begin{equation}\label{spkswa}
{\cal I}m\lambda_{\psi K}=0.734\pm0.054.
\end{equation}
(The world average given in eq. (\ref{spkswa}) includes the new
results from Belle \cite{yamauchi}, ${\cal I}m\lambda_{\psi
  K}=0.719\pm0.074\pm0.035$, and Babar \cite{karyotakis}, ${\cal I}m\lambda_{\psi
  K}=0.741\pm0.067\pm0.033$, and previous results from Opal, Aleph and CDF.)

In addition, CP asymmetries in many other modes (neutral $B$ decays
into final CP eigenstates and non-CP eigenstates and charged $B$
decays) have been searched for. We describe the implications of
the new data for our theoretical understanding of CP violation.

\section{Standard Model Lessons}

Within the Standard Model, the only source of CP violation is the
Kobayashi-Maskawa (KM) phase \cite{Kobayashi:fv}. This phase appears
in the CKM matrix which describes the charged current interactions of
quarks.

\subsection{Unitarity Triangles}
\begin{figure*}[htb]
  \includegraphics*[width=0.5\textwidth]{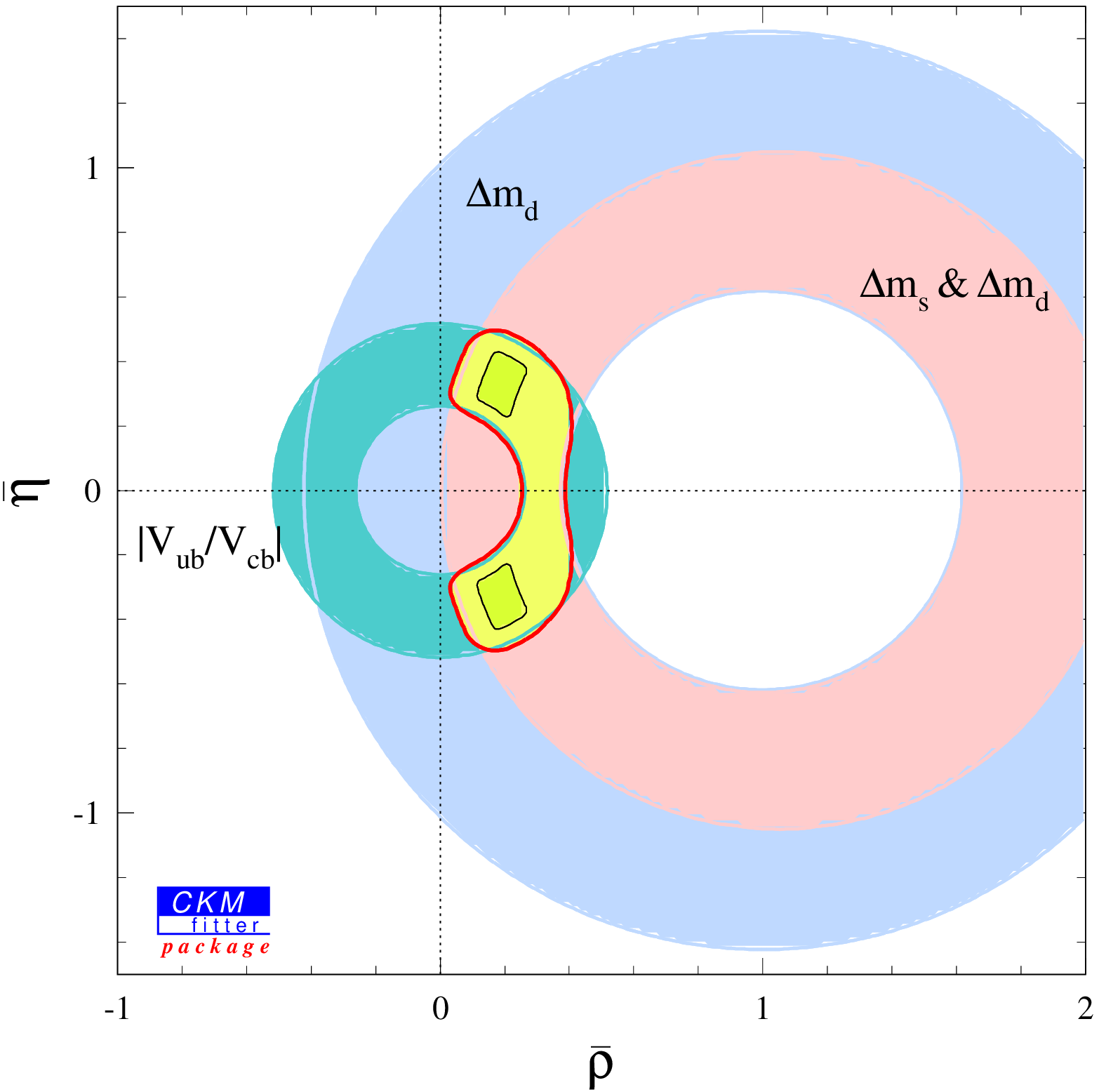}\hfill
  \includegraphics*[width=0.5\textwidth]{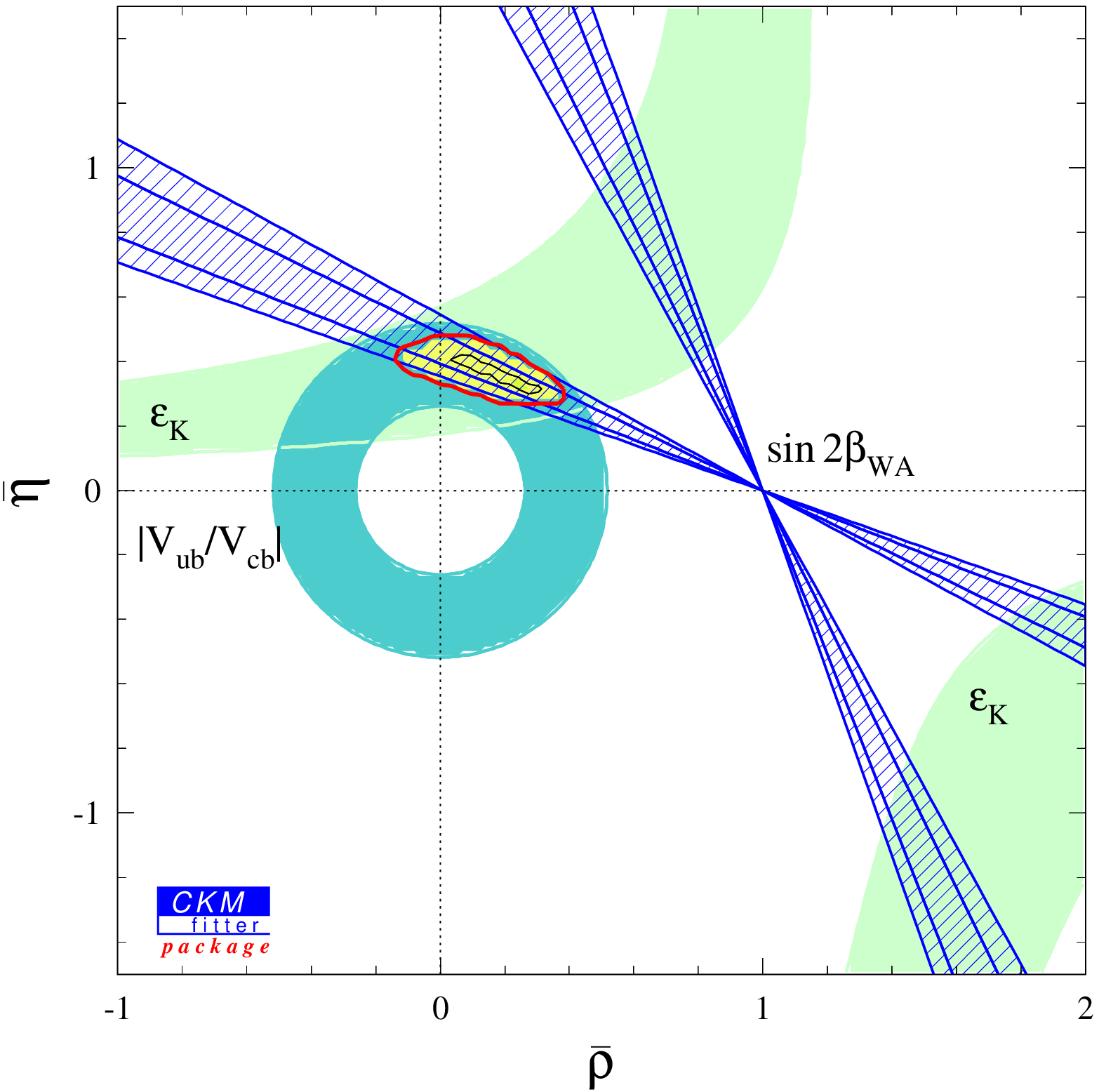}
\caption{Constraints on the $(\bar\rho,\bar\eta)$ parameters from tree
  processes and  from (left) CP conserving loop processes ($\Delta m_B,\
  \Delta m_{B_s}$) and (right) CP violating processes ($\varepsilon,\
  {\cal I}m\lambda_{\psi K}$).}
\label{fig:cpcvut}
\end{figure*}
\begin{figure*}[htb]
  \includegraphics*[width=0.5\textwidth]{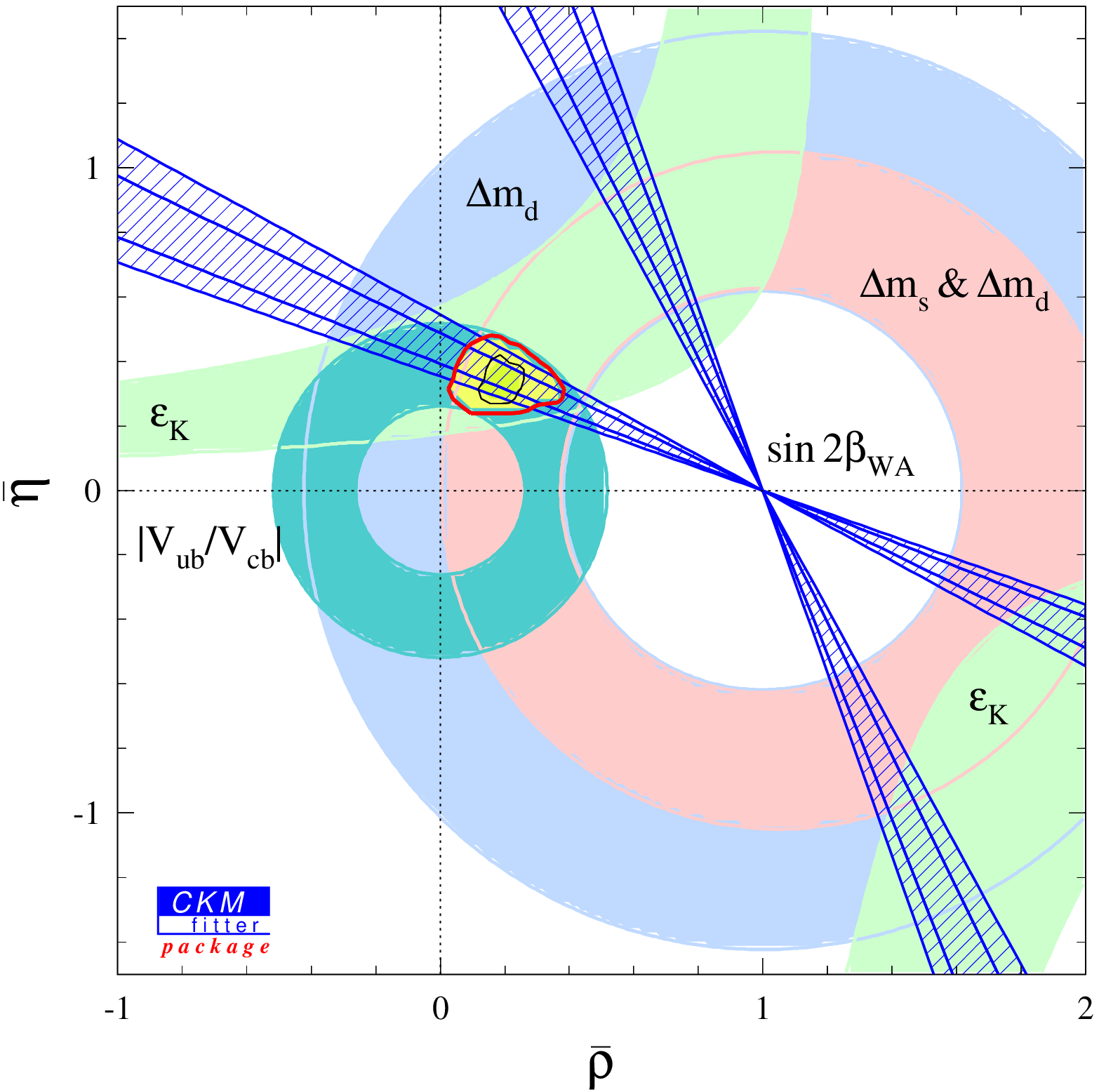}\hfill
  \includegraphics*[width=0.5\textwidth]{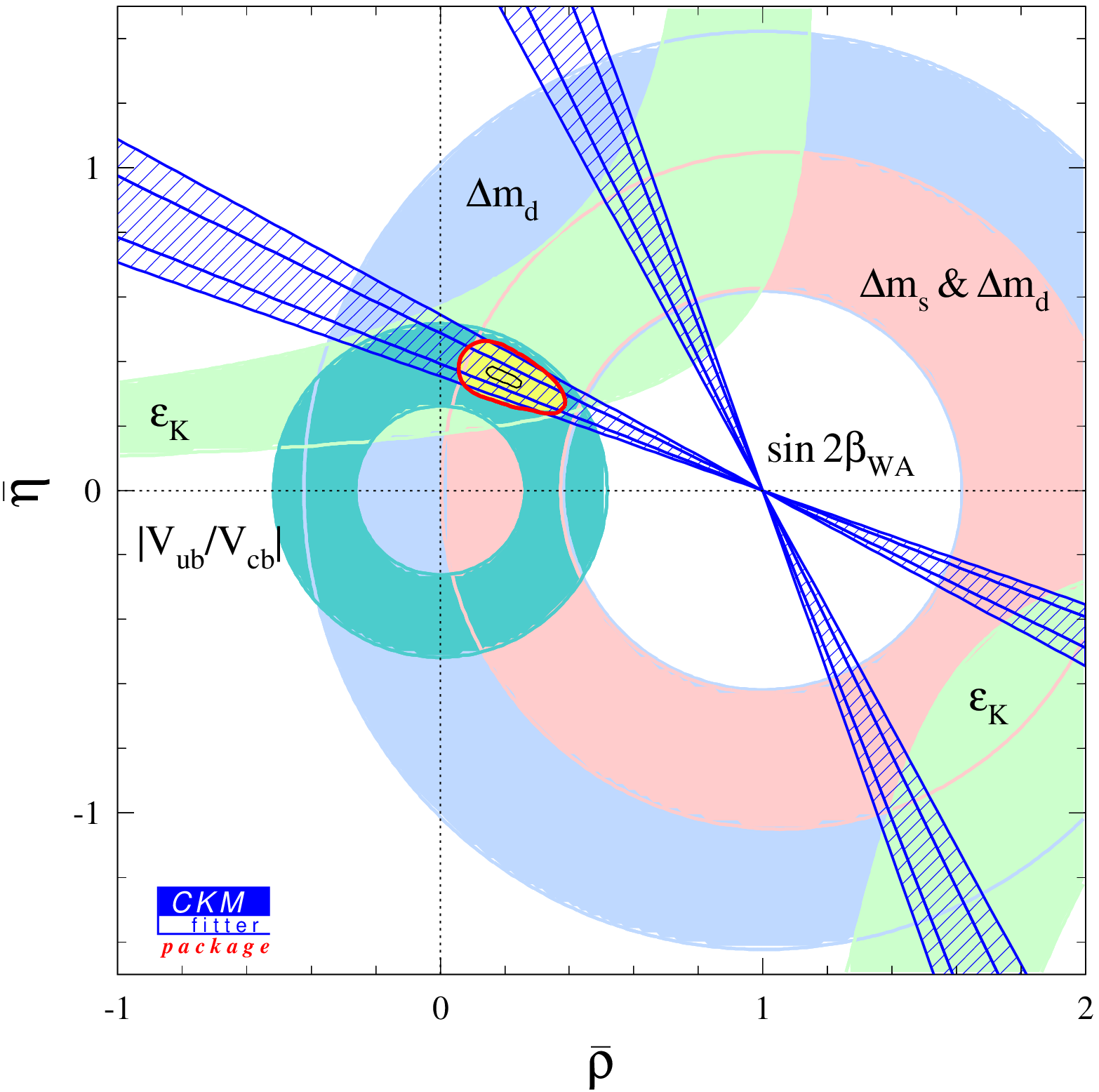}
  \caption{Constraints on the $(\bar\rho,\bar\eta)$ parameters from
    (left) CP conserving and the $\varepsilon$ observables compared to
    the ${\cal I}m\lambda_{\psi K}$ constraint, and (right) from all
    observables.} 
\label{fig:smut}
\end{figure*}
\begin{figure*}[htb]
  \includegraphics*[width=0.5\textwidth]{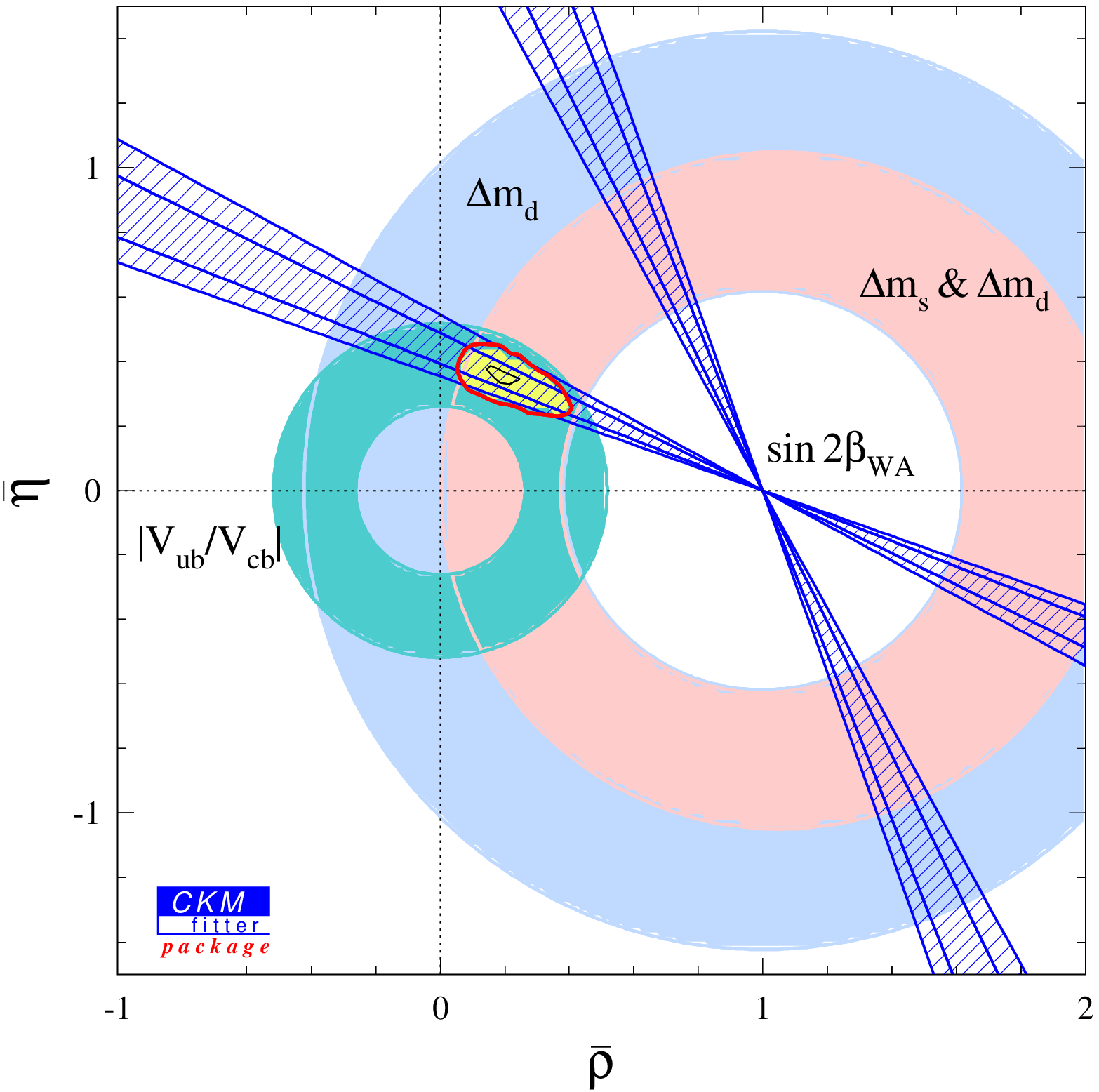}\hfill
  \includegraphics*[width=0.5\textwidth]{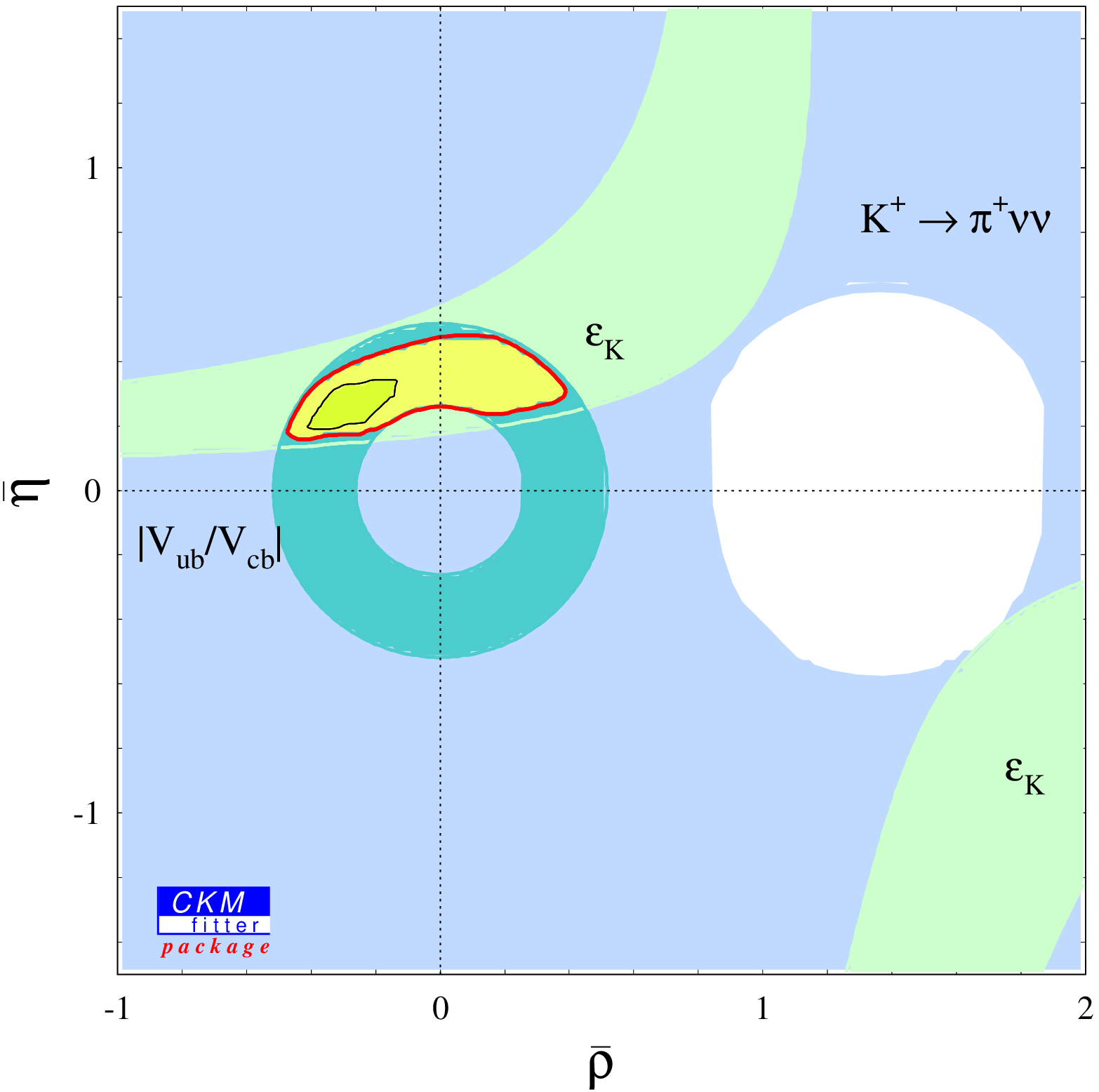}
  \caption{Constraints on the $(\bar\rho,\bar\eta)$ parameters from
    (left) $B$ physics ($\Delta m_B,\ \Delta m_{B_s},\
{\cal I}m\lambda_{\psi K}$), and (right) $K$ physics ($\varepsilon,\
{\cal B}(K^+\to\pi^+\nu\bar\nu)$).} 
\label{fig:bkut}
\end{figure*}

The CKM matrix gives the couplings of the $W^+$-boson to $\bar u_i
d_j$ quark pairs:
\begin{equation}\label{ckmvij}
  V=   \pmatrix{
    V_{ud} & V_{us} & V_{ub} \cr
    V_{cd} & V_{cs} & V_{cb} \cr
    V_{td} & V_{ts} & V_{tb} \cr}.
\end{equation}
The unitarity of the matrix leads to various relations among its
elements, {\it e.g.},
\begin{equation}\label{unirel}
  V_{ud}V_{ub}^*+V_{cd}V_{cb}^*+V_{td}V_{tb}^*=0.
\end{equation}
The {\it unitarity triangle} is a geometrical presentation in the
complex plane of the relation (\ref{unirel}). It provides a convenient
tool in the study of flavor physics and CP violation:

1. Flavor changing processes and unitarity give rather
precise information on the magnitudes of all elements except for
$|V_{ub}|$ and $|V_{td}|$, which have large uncertainties. The
unitarity triangle is a pictorial way of combining the various
constraints on these elements and of testing whether the
constraints can be explained consistently within the CKM framework.

2. The angles of the triangle are related to CP violation. In
  particular, measurements of various CP asymmetries in $B$ decays can
  be used to constrain the values of these angles. Conversely, the
  consistency of the various constraints tests whether CP violation can
  be accounted for by the Kobayashi-Maskawa mechanism.
  
Since the length of one side, $|V_{cd}V_{cb}|$, is well known,
it is convenient to re-scale the unitarity triangle by the length of
this side and put it on the real axis. When doing so, the coordinates
of the remaining vertex correspond to the $\rho$ and $\eta$ parameters
in the Wolfenstein parametrization \cite{Wolfenstein:1983yz} of the
CKM matrix (or, in an improved version \cite{Buras:1994ec}, to
$\bar\rho=(1-{\lambda^2\over2})\rho$ and
$\bar\eta=(1-{\lambda^2\over2})\eta$): 
\begin{equation}\label{ckmvij}
    V=\pmatrix{
       1-{\lambda^2\over2}&\lambda&A\lambda^3(\rho-i\eta)\cr
       -\lambda&1-{\lambda^2\over2}&A\lambda^2\cr
       A\lambda^3(1-\rho-i\eta)&-A\lambda^2&1\cr}.
   \end{equation}

\begin{figure}[htb]
  \includegraphics*[angle=90,width=\columnwidth]{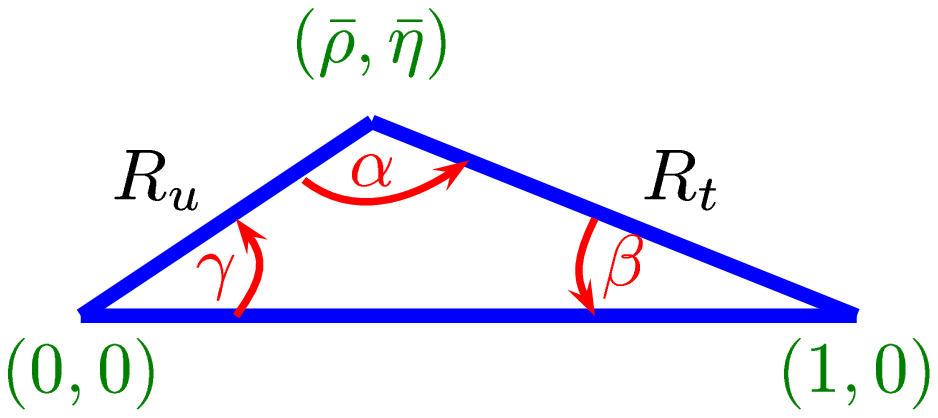}
\end{figure}

The angles $\alpha,\beta$ and $\gamma$ (also known as, respectively,
$\phi_2,\phi_1$ and $\phi_3$) are defined as follows:
\begin{eqnarray}\label{defabc}
  \alpha&\equiv&\arg\left(-{V_{td}V_{tb}^*\over V_{ud}V_{ub}^*}\right),\
  \ \  \beta\equiv\arg\left(-{V_{cd}V_{cb}^*\over
  V_{td}V_{tb}^*}\right),\nonumber\\ 
  \gamma&\equiv&\arg\left(-{V_{ud}V_{ub}^*\over
  V_{cd}V_{cb}^*}\right).
\end{eqnarray}
The lengths $R_t$ and $R_u$ are defined as follows:
\begin{equation}\label{defabc}
  R_t\equiv\left|{V_{td}V_{tb}^*\over V_{ud}V_{ub}^*}\right|,\
  \ \ \ \  R_u\equiv\left|{V_{ud}V_{ub}^*\over
  V_{cd}V_{cb}^*}\right|.
\end{equation}
In what follows, we present the constraints on the
$(\bar\rho,\bar\eta)$ parameters coming from various classes of
processes. (The plots have been produced using the CKMFitter package
\cite{Hocker:2001xe}.) 

In Figure \ref{fig:cpcvut} we compare the constraints from CP conserving
processes to those from CP violating ones. The CP conserving observables
are the mass difference in the neutral $B$ system, $\Delta m_B$,
and the lower bound on the mass difference in the $B_s$ system,
$\Delta m_{B_s}$.  The CP violating observables are the indirect CP
violation in $K\to\pi\pi$ decays, $\varepsilon$, and the CP asymmetry in 
$B\to\psi K_S$ decays, ${\cal I}m\lambda_{\psi K}$.
There are two important lessons to be drawn from this comparison: 
\begin{itemize}
  \item Since there is a significant overlap between the allowed
    regions in the two panels, we learn that the two sets of
    constraints are consistent with each other. Thus it is very likely
    that the KM mechanism is indeed the source of the observed CP
    violation.
    \item The constraints from the CP violating observables are
      stronger than those from the CP conserving ones. While the
      allowed ranges are related to the experimental accuracy, an
      important factor in this situation is the fact that CP is a good
      symmetry of the strong interactions. (The effects of
      $\theta_{\rm QCD}$ are irrelevant to meson decays.)
      Consequently, some CP asymmetries can be theoretically
      interpreted with practically zero hadronic uncertainties.
    \end{itemize}

Another way to see the consistency of the KM picture of CP violation
is the following. Within the Standard Model, there is a single CP
violating parameter. Therefore, roughly speaking, a measurement of a
single CP violating observable simply determines the value of this
parameter. This situation is demonstrated in the left panel of Figure
\ref{fig:smut}, where the constraints from all but the ${\cal
  I}m\lambda_{\psi K}$-measurement are used to produce an allowed
range in the $(\bar\rho,\bar\eta)$ plane. A second measurement of a CP
violating observable tests this mechanism, as demonstrated in the
same Figure by overlaying the constraint from the measurement of
${\cal I}m\lambda_{\psi K}$. (It is amusing to note that in
ref. \cite{stocchi}, the allowed range for ${\cal I}m\lambda_{\psi K}$
from the fit to all other observables is quoted to be
$0.734^{+0.055}_{-0.045}$, to be compared with the range from the
direct measurements in eq. (\ref{spkswa}).) The allowed region in the
($\bar\rho,\bar\eta$) plane from the combination of all observables is
shown in the right panel of Figure \ref{fig:smut}.
We can again draw several conclusions:

\begin{itemize}
  \item The CKM matrix provides a consistent picture of all the
    measured flavor and  CP violating processes.
    \item The recent measurement of ${\cal I}m\lambda_{\psi K}$ adds a
      significant constraint.
    \end{itemize}

In Figure \ref{fig:bkut} we make one final comparison, between
observables related to $B$ physics ($\Delta m_B,\ \Delta m_{B_s},\
{\cal I}m\lambda_{\psi K}$, left panel) and to $K$ physics
($\varepsilon,\ {\cal B}(K^+\to\pi^+\nu\bar\nu)$ \cite{Adler:2001xv},
right panel). 

The conclusions that we draw from this comparison are the following:
\begin{itemize}
  \item There is no signal of new flavor physics.
    \item At present, the constraints from $B$ physics are much
      stronger. Future measurements of ${\cal B}(K\to\pi\nu\bar\nu)$
      (for both the charged and the neutral modes) will be essential
      to make this comparison into a useful probe of new physics.
    \end{itemize}

\subsection{Testing the KM mechanism}
Since, by the consistency between the predicted range and the measured
value of the CP asymmetry in $B\to\psi K_S$, the KM mechanism of CP
violation has successfully passed its first precision test, we are
able to make the following statement:

{\it Very likely, the KM mechanism is the dominant source of CP
  violation in flavor changing processes.}

Thirty eight years have passed since the discovery of CP violation
\cite{Christenson:fg} and twenty nine years have passed since the KM
mechanism has been proposed \cite{Kobayashi:fv}. But only now,
following the impressively precise measurements by Belle and Babar
that yield (\ref{spkswa}), we can make the above statement based on
experimental evidence. This is a very important step forward in our
theoretical understanding of CP violation.

We would like to emphasize, however, three important points in the
above statement:
\begin{enumerate}
  \item {\it `Very likely:'} since we are using only two CP violating
    observables, the consistency of the KM picture could be
    accidental. Additional measurements are crucial to make a more
    convincing case for the validity of the KM mechanism.
    \item {\it `Dominant:'} the accuracy of the Standard Model
      prediction for ${\cal I}m\lambda_{\psi K}$ is of ${\cal
        O}(20\%)$. Therefore, it is quite possible that new physics
      contributes to meson decays at this level.
      \item {\it `Flavor changing:'} while we have good reasons to
        think that flavor changing CP violation is dominated by
        Standard Model processes, the situation could be very
        different for flavor diagonal CP violation. Here, the KM
        mechanism predicts unobservably small CP violation, while new
        physics can dominate such observables by several orders of
        magnitude. Future searches of EDMs are crucial to clarify this
        point.
      \end{enumerate}

\section{General Lessons}

CP violation in meson decays is a complex phenomenon. This is best
demonstrated by examining CP violation in neutral meson decay into a
final CP eigenstate. ($P$ stands here for any of the $K$, $D$, $B$ and
$B_s$ mesons.)

\begin{figure}[htb]
  \includegraphics*[width=\columnwidth]{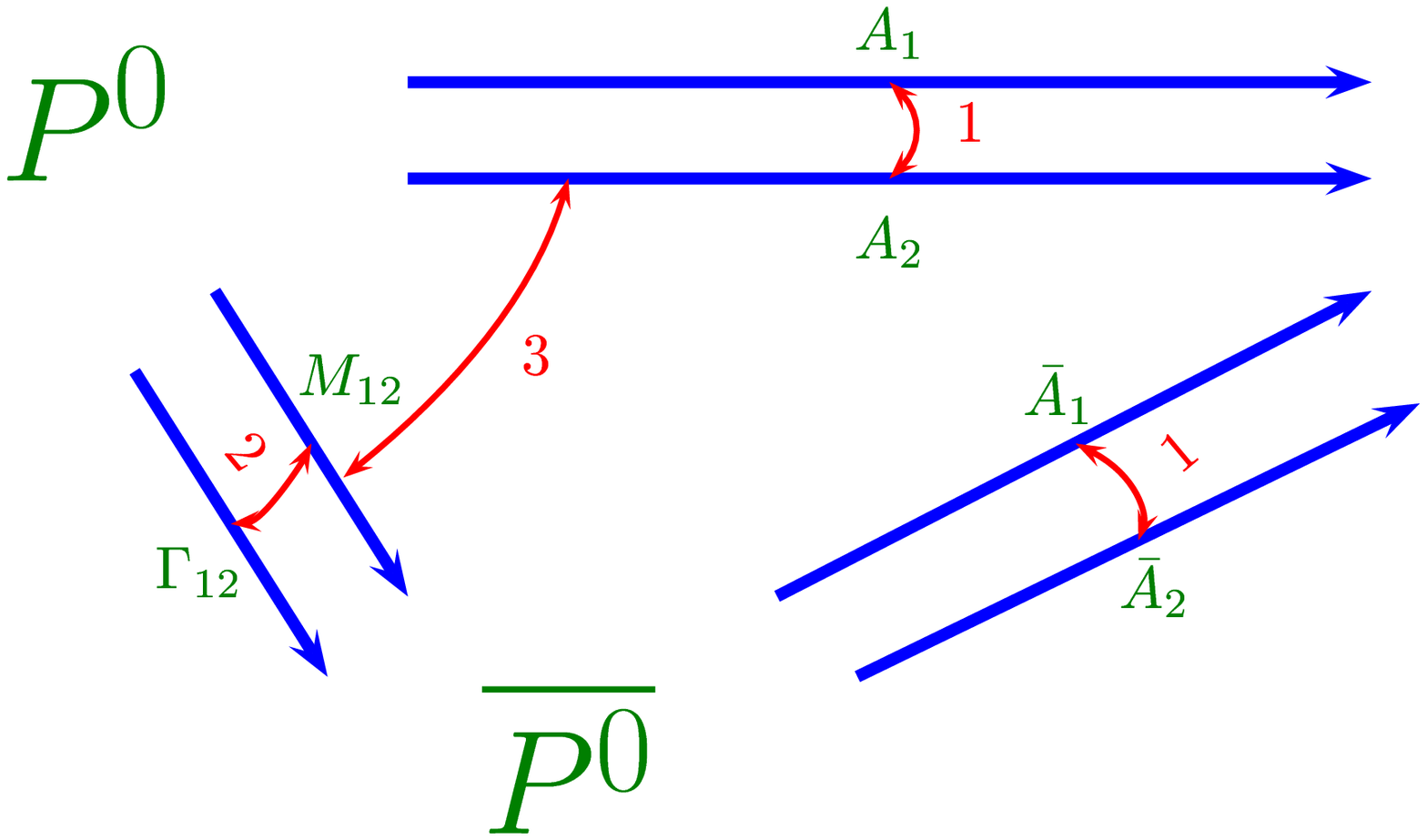}
\label{fig:btofcp}
\end{figure}

Each arrow in this figure stands for an amplitude that carries an
independent CP violating phase. The various interferences between the
different paths from $P^0$ to $f_{\rm CP}$ yield three distinct
manifestations of CP violation:

\begin{enumerate}
\item {\it In decay:}
  \begin{equation}\label{indec}
    \left|{\bar A\over A}\right|\neq1\ \ \
    \left[{\bar A\over A}={\bar A_1+\bar A_2\over A_1+A_2}\right].
  \end{equation}
\item {\it In mixing:}
  \begin{equation}\label{inmix}
  \left|{q\over p}\right|\neq1\ \ \  
\left[\left({q\over
      p}\right)^2={2M_{12}^*-i\Gamma_{12}^*\over2M_{12}-i\Gamma_{12}}\right].
\end{equation}
\item {\it In interference (of decays with and without mixing):}
  \begin{equation}\label{inint}
{\cal I}m\lambda\neq0\ \ \ 
\left[\lambda={q\over p}{\bar A\over A}\right].
\end{equation}
\end{enumerate}

One of the beautiful features of CP violation is that experiments can
measure each of these three types separately. Take for example $B$
meson decays:

1. The CP asymmetry in charged $B$ decays is
  sensitive to only CP violation in decay:
  \begin{eqnarray}\label{acpch}
  {\cal A}_{f^\mp}&\equiv&{\Gamma(B^-\to f^-)-\Gamma(B^+\to
     f^+)\over\Gamma(B^-\to f^-)+\Gamma(B^+\to f^+)}\nonumber\\ 
   &=&{|\bar A_f/A_f|^2-1\over|\bar A_f/A_f|^2+1}.
 \end{eqnarray}

2. The CP asymmetry in semileptonic neutral $B$ decays is
  sensitive to only CP violation in mixing:
  \begin{eqnarray}\label{acpsl}
 {\cal A}_{\rm SL}&\equiv&{\Gamma(\bar B^0_{\rm
     phys}\to\ell^+X)-\Gamma(B^0_{\rm phys}\to\ell^-X)
     \over\Gamma(\bar B^0_{\rm phys}\to\ell^+X)+\Gamma(B^0_{\rm
     phys}\to\ell^-X)}\nonumber\\ 
   &=&{1-|q/p|^4\over1+|q/p|^4}.
 \end{eqnarray}

3. The CP asymmetry in neutral $B$ decays probes separately 
  $|\lambda|$ (a combination of CP violation in mixing and in decay)
  and ${\cal I}m\lambda$ (purely CP violation in the interference of
  decays with and without mixing)
  \cite{Dunietz:1986vi,Blinov:ru,Bigi:1986vr}:    
   \begin{eqnarray}\label{acpcp}   
 {\cal A}_{f_{\rm CP}}(t)&\equiv&
{\Gamma(\bar B^0_{\rm phys}\to f_{\rm CP})-\Gamma(B^0_{\rm phys}\to f_{\rm CP})
     \over\Gamma(\bar B^0_{\rm phys}\to f_{\rm CP} )+\Gamma(B^0_{\rm
       phys}\to f_{\rm CP} )}\nonumber\\
  &=&-C_{f_{\rm CP}}\cos(\Delta
      m_Bt)+S_{f_{\rm CP}}\sin(\Delta m_Bt)\nonumber\\  
      C_{f_{\rm CP}}&=&-{\cal A}_{f_{\rm
   CP}}={1-|\lambda_{f_{\rm CP}}|^2\over 1+|\lambda_{f_{\rm
   CP}}|^2},\nonumber\\
          S_{f_{\rm CP}}&=&{2{\cal I}m\lambda_{f_{\rm
                CP}}\over1+|\lambda_{f_{\rm CP}}|^2}.
        \end{eqnarray}
        
Note that Babar's $C_{f_{\rm CP}}$ corresponds to Belle's $-{\cal
A}_{f_{\rm CP}}$. Further note that the latter notation suggests that this CP
asymmetry is analogous to ${\cal A}_{f^\mp}$ measured in charged $B$
decays. Formally, ${\cal A}_{f_{\rm CP}}$ measures deviations of
$|\lambda|$ from unity while ${\cal A}_{f^\mp}$ measures deviations of
$|\bar A/A|$ from one. However, the deviation of $|q/p|$ from unity is
known to be $\lsim{\cal O}(10^{-2})$ and, given the present
sensitivity of searches for ${\cal A}_{f_{\rm CP}}\neq0$, can be
safely neglected. In practice, therefore, the analogy is justified. 

There is a class of CP asymmetries in neutral meson decays into final
CP eigenstates that theorists love most. It involves decays where the
direct decay amplitude is dominated by a single weak phase, so that CP
violation in decay can be neglected ($|\bar A/A|=1$) and where the
effect of CP violation in mixing can be neglected ($|q/p|=1$). In this
case $|\lambda|=1$, and the only remaining CP violating effect is
$S_{f_{\rm CP}}\neq0$. 

\begin{figure}[htb]
  \includegraphics*[angle=90,width=\columnwidth]{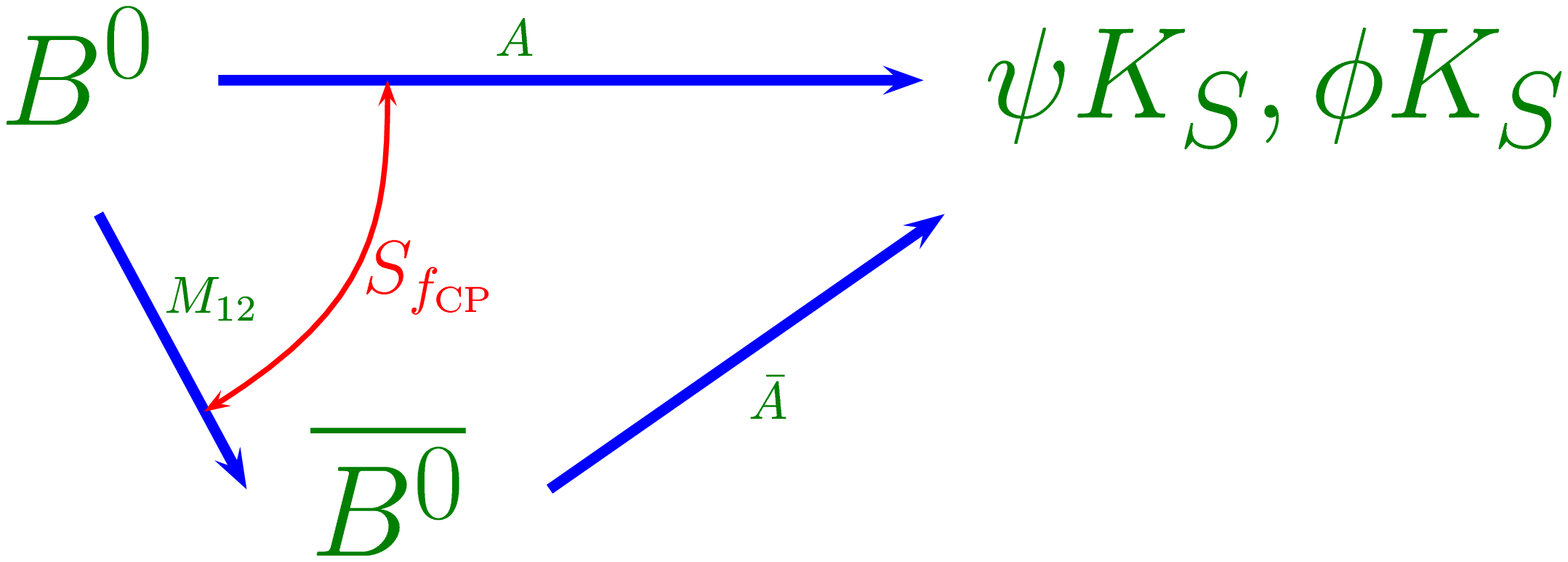}
\label{fig:btofcp1}
\end{figure}

The reason that this case is the theorists' favorite is that the
theoretical interpretation, in terms of Lagrangian parameters, is
uniquely clean. Explicitly, the asymmetry can be expressed purely in
terms of the (CP violating) phase difference between the mixing
amplitude and twice the decay amplitude:
\begin{equation}\label{cleanasy}
  S_{f_{\rm CP}}={\cal I}m\lambda_{f_{\rm
  CP}}=\pm\sin[\arg(M_{12}^*)-2\arg(A_{f_{\rm CP}})],
\end{equation}
where the sign depends on the CP eigenvalue of the final state. Among
the few modes that belong to this class are the $B\to\psi K_S$,
$B\to\phi K_S$ \cite{Carter:hr,Bigi:qs} and $K\to\pi\nu\bar\nu$ 
\cite{Littenberg:ix} decays.

\subsection{$K\to\pi\pi$}
All three types of CP violation have been measured in $K\to\pi\pi$
decays. Historically, a different language has been used to
parametrize the various CP violating observables. The translation
between that language and our language is straightforward:
\begin{eqnarray}\label{epsepsp}
  \varepsilon&=&{1-\lambda_0\over1+\lambda_0},\nonumber\\
  \varepsilon^\prime&=&{1\over6}(\lambda_{00}-\lambda_{+-}),
\end{eqnarray}
where the subindex $0$ refers to final two pions in an isospin-zero
state, while the $+-$ and $00$ sub-indices refer to, respectively,
$\pi^+\pi^-$ and $\pi^0\pi^0$ final states.

\begin{figure}[htb]
  \includegraphics*[angle=270,width=\columnwidth]{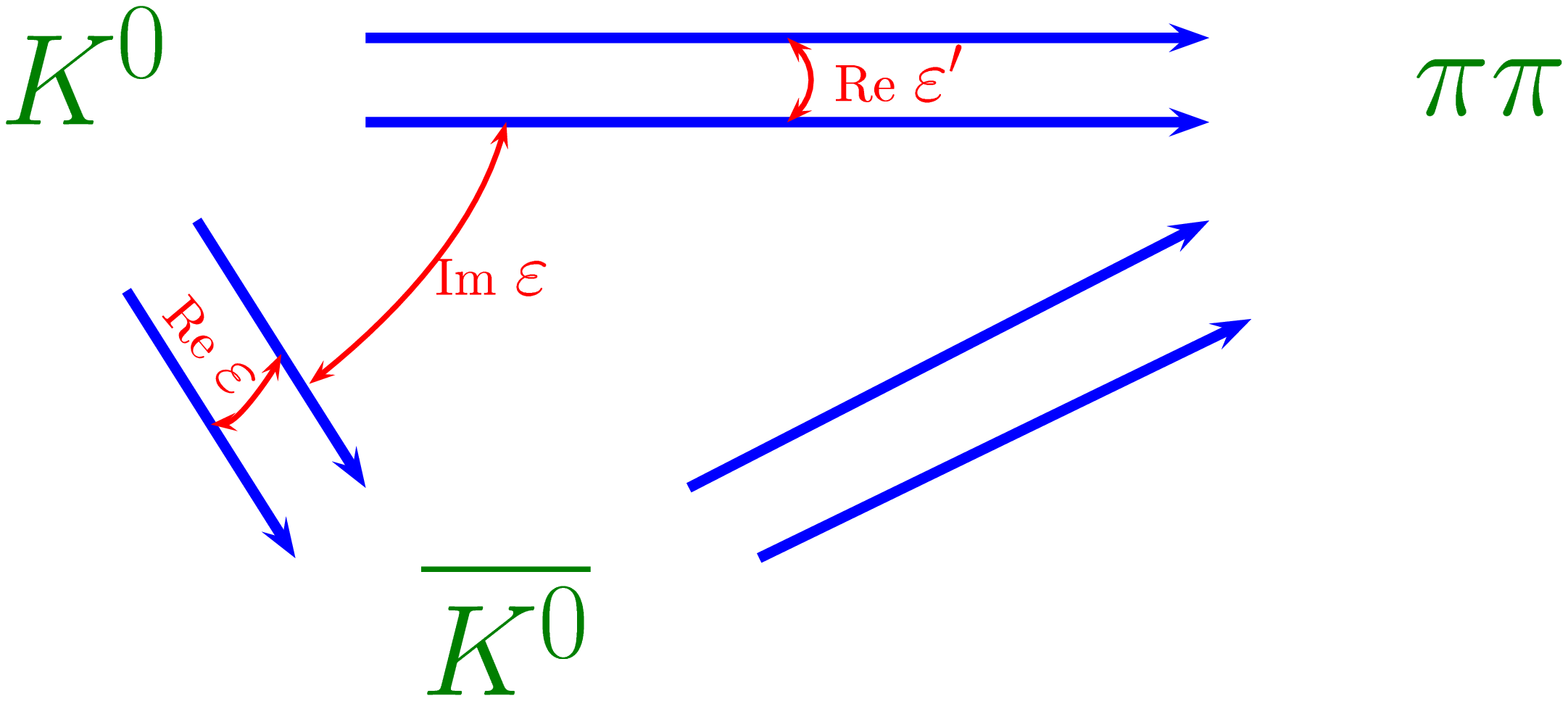}
\label{fig:ktopipi}
\end{figure}

The experimental values of the $\varepsilon$ and $\varepsilon^\prime$
parameters are given in eqs. (\ref{epskwa}) and (\ref{epspwa}). The
measurement of $\varepsilon$ in 1964 constituted the discovery of CP
violation and drove, through the work of Kobayashi and Maskawa, to the
prediction that a third generation exists. The precise measurement of
${\cal R}e(\varepsilon^\prime/\varepsilon)$ has important
implications: 
\begin{itemize}
  \item Direct CP violation has been observed.
  \item The superweak scenario \cite{Wolfenstein:ks} is excluded.
  \item The result is consistent with the SM predictions.
  \item Large hadronic uncertainties make it impossible to extract a
    useful CKM constraint. This is the reason that no
    $\varepsilon^\prime$ constraint appears in our unitarity
    triangles.  
  \item New physics ({\it e.g.} Supersymmetry \cite{Masiero:1999ub})
    may contribute significantly.
    \end{itemize}

\subsection{$B\to\psi K_S$}
The $B\to\psi K_S$ decay \cite{Carter:hr,Bigi:qs}, related to the
$\bar b\to\bar cc\bar s$ 
quark transition, is dominated by a single weak
phase. An amplitude carrying a second, different phase, is suppressed
by both a CKM factor of ${\cal O}(\lambda^2)$ and a loop
factor. Consequently, the deviation of $|\bar A_{\psi K}/A_{\psi K}|$
from unity is predicted to be below the percent level. It is then
expected that $|\lambda_{\psi K}|=1$ to an excellent
approximation. This golden mode belongs then to the `clean' class
described above: $S_{\psi K}={\cal I}m\lambda_{\psi K}$ and can be
cleanly interpreted in terms of the difference between the phase
of the $B^0-\overline{B^0}$ mixing amplitude and twice the phase of
the $b\to c\bar cs$ decay amplitude.

It would be nice to confirm this expectation in a model independent
way. The present average over the Babar \cite{karyotakis} and Belle
\cite{yamauchi} measurements is
\begin{equation}\label{cpsiks}
  |\lambda_{\psi K}|=\left|{q\over p}{\bar
      A_{\psi K}\over A_{\psi K}}\right|=0.949\pm0.039.
\end{equation}
The central value is five percent away from unity with an error of
order four percent. While this result is certainly consistent with
$|\lambda_{\psi K}|=1$, it does not yet confirm it. We would like to
argue that the information from two other, entirely independent
measurements constrains the deviation of $|\lambda_{\psi K}|$ from one
to a much better accuracy. The first measurement is that of the CP
asymmetry in semileptonic $B$ decays. The world average over the
results from Opal, Cleo, Aleph and Babar (see \cite{Laplace:2002ik}
and references therein) is given by
\begin{equation}\label{aslexp}
  {\cal A}_{\rm SL}=0.002\pm0.014.
\end{equation}
Through eq. (\ref{acpsl}) we can constrain the size of CP violation in
mixing:
\begin{equation}\label{qpexp}
  |q/p|=0.999\pm0.007.
\end{equation}
The second measurement is that of the CP asymmetry in the charged
mode, $B^\mp\to\psi K^\mp$:
\begin{equation}\label{achexp}
 {\cal A}_{\psi K^\mp}=0.008\pm0.025.
\end{equation}
(This is the average over Cleo \cite{Bonvicini:2000dr} and Babar
\cite{Aubert:2002yv} measurements.) Through eq. (\ref{acpch}), and
using isospin symmetry to relate this measurement to the neutral mode
\cite{Fleischer:2001cw}, we can constrain the size of CP violation in
decay: 
\begin{equation}\label{aaexp}
 |\bar A_{\psi K}/A_{\psi K}|=1.008\pm0.025.
\end{equation}
Together, eqs. (\ref{qpexp}) and (\ref{aaexp}) give
\begin{equation}
  |\lambda_{\psi K}|=1.007\pm0.026.
\end{equation}
This is a much stronger constraint than (\ref{cpsiks}) and allows us
to confidently use $|\lambda_{\psi K}|=1$ from here on. This exercise
carries an important lesson: the two measurements employed here
provide only upper bound on asymmetries that are not subject to a
clean theoretical interpretation. Yet, they take us a step forward in
our understanding of CP violation. Indeed, each and every measurement
contributes its part to solving the big puzzle of CP violation.

What do we learn from the measurement of ${\cal I}m\lambda_{\psi
  K}=0.734\pm0.054$?

$\bullet$ CP violation has been observed in $B$ decays.

$\bullet$ The Kobayashi-Maskawa mechanism of CP violation has
  successfully passed its first precision test. Figure \ref{fig:smut}
  makes a convincing case for this statement.

$\bullet$  A significant constraint on the CKM parameters
($\bar\rho,\bar\eta$) has been added. Within the Standard Model,
\begin{equation}\label{lampks}
  \lambda_{\psi K_{S,L}}=\mp\left({V_{tb}^*V_{td}\over
      V_{tb}V_{td}^*}\right)\left({V_{cb}V_{cd}^*\over
      V_{cb}^*V_{cd}}\right)=\mp e^{-2i\beta}.
\end{equation}
The first CKM factor in (\ref{lampks}) comes from $(q/p)$ and the
second from $\bar A_{\psi K}/A_{\psi K}$. The $\mp$ signs are related
to the CP eigenvalue of the final state. One obtains
\cite{Dunietz:1986vi,Bigi:1986vr,Dib:1989uz}
  \begin{equation}\label{apksstb}
   S_{\psi K_S}= {\cal I}m\lambda_{\psi
      K_S}=\sin2\beta={2\bar\eta(1-\bar\rho)\over\bar\eta^2+(1-\bar\rho)^2}.
  \end{equation}
This is the constraint that has been used in our unitarity
triangles. Note that there are no hadronic parameters involved in the
translation of the experimental value of $S_{\psi K_S}$ to an allowed
region in the ($\bar\rho,\bar\eta$) plane. Hadronic uncertainties
arise only below the percent level.

Our CKM fit yields, for example, the following allowed ranges (at
CL$>32\%$):
\begin{eqnarray}\label{fitrhoeta}
  0.12\leq\bar\rho\leq0.35,&\ \ 
    &0.28\leq\bar\eta\leq0.41,\nonumber\\
    -0.82\leq\sin2\alpha\leq0.24,&\ \ \ &40^o\leq\gamma\leq73^o.
  \end{eqnarray}
  
$\bullet$ Approximate CP is excluded. Approximate CP has been one of the
more intriguing alternatives to the KM mechanism. It assumes that
$\varepsilon$ and $\varepsilon^\prime$ are small not because of flavor
suppression, as is the case with the KM mechanism, but because all CP
violating phases are small. This idea can be realized (and is well
motivated) within the supersymmetric framework
\cite{Babu:1993ai,Abel:1996eb,Eyal:1998bk,Babu:1999xf,Eyal:1999gk}.
However, the observation of a CP asymmetry of order one excludes this
idea. (One can still write viable models of approximate CP, but these
involve fine-tuning.) Similarly, minimal left-right-symmetric models
with spontaneous CP violation
\cite{Ball:1999mb,Ball:1999yi,Bergmann:2001pm} are excluded.

$\bullet$  New, CP violating physics that contributes $>20\%$ to
  $B^0-\overline{B^0}$ mixing is disfavored. As mentioned above, it is
  still possible that there is a significant new physics in
  $B^0-\overline{B^0}$ mixing, but the new phase and the Standard
  Model $\beta$ (different from the Standard Model fit) conspire to
  give an asymmetry that is the same as predicted by the Standard
  Model. This situation is rather unlikely, but is not rigorously
  excluded. 

\subsection{$B\to\phi K_S$ and $B\to\eta^\prime K_S$}
The $B\to\phi K_S$ and $B\to\eta^\prime K_S$ decays, related to the
$\bar b\to\bar ss\bar s$ quark transition, are dominated by a single weak
phase. An amplitude carrying a second, different phase, is suppressed
by a CKM factor of ${\cal O}(\lambda^2)$. Consequently, the deviation
of $|\bar A/A|$ from unity is predicted to be at the few ($\lsim4$) percent
level \cite{Grossman:1997gr,London:1997zk}. It is then expected that
$|\lambda_{\phi K}|=|\lambda_{\eta^\prime K}|=1$ to a good
approximation. These modes belong then to the `clean' class
described above: $S\simeq{\cal I}m\lambda$ and can be
cleanly interpreted in terms of the difference between the phase
of the $B^0-\overline{B^0}$ mixing amplitude and twice the phase of
the $b\to s\bar ss$ decay amplitude.  

Averaging over the new Belle \cite{yamauchi} and Babar
\cite{karyotakis} results for the $\phi K_S$ mode, we obtain the
present world average, 
\begin{eqnarray}\label{phiksexp}
  C_{\phi K_S}&=&+0.56\pm0.43,\nonumber\\
  S_{\phi K_S}&=&-0.39\pm0.41.
  \end{eqnarray}
The new Belle results for the $\eta^\prime K_S$ mode
\cite{yamauchi} read
\begin{eqnarray}\label{etaksexp}
  C_{\eta^\prime K_S}&=&-0.26\pm0.22,\nonumber\\
  S_{\eta^\prime K_S}&=&+0.76\pm0.36.
\end{eqnarray}
Thus, CP violation in $b\to s\bar ss$ transitions has not yet been
observed.

In spite of being related to a different quark transition,
the CKM dependence of these two modes is the same [up to ${\cal
  O}(\lambda^2)$ effects] as that of $\lambda_{\psi K}$ of
eq. (\ref{lampks}). Consequently, within the Standard Model,
\begin{equation}\label{phipsi}
  S_{\phi K_S}\simeq S_{\eta^\prime K_S}\simeq S_{\psi K_S}.
\end{equation}
A difference (larger than a few percent) between the CP asymmetries in
the $\phi K_S$ or $\eta^\prime K_S$ modes and in the $\psi K_S$ mode is a
clear signal of new physics \cite{Grossman:1996ke}. More specifically,
such a difference requires that new, CP violating physics contributes
significantly to $b\to s$ transitions.

The measurements of the two modes suffer from large statistical
errors. At present, there is no evidence ({\it i.e.}, an effect
$\geq3\sigma$) for either  $S_{\phi K_S}\neq S_{\psi K_S}$ or
$S_{\eta^\prime K_S}\neq S_{\psi K_S}$. We conclude that, at present,
there is no evidence for new physics in these measurements.

One might be tempted to interpret the $2.7\sigma$ difference between
$S_{\phi K_S}$ and $S_{\psi K_S}$ as a {\it hint} for new physics. It
would be rather puzzling, however, (though, perhaps, not impossible)
if new, CP violating physics affects $B\to\phi K_S$ in a dramatic way
but gives only a very small effect in $B\to\eta^\prime
K_S$. Furthermore, Belle has also searched for CP violation in $B\to
K^+K^- K_S$ decays (with the $\phi$-resonance contributions removed)
\cite{yamauchi}:
\begin{eqnarray}\label{kkkexp}
  C_{K^+K^-K_S}&=&+0.42\pm0.37^{+0.22}_{-0.03},\nonumber\\
  S_{K^+K^-K_S}&=&-0.52\pm0.47^{+0.03}_{-0.27}.
\end{eqnarray}
(This is a-priori not a CP eigenstate, but a combination of
experimental data and isospin considerations allows Belle to conclude
that the final state is dominantly CP-even. The last, asymmetric,
error is related to the fractions of CP-even and CP-odd components.)
We learn that there is no observed deviation from 
$-S_{K^+K^-K_S}=S_{\psi K_S}$. We would like to suggest then that,
when speculating on the source of the difference between $S_{\phi
  K_S}$ and $S_{\psi K_S}$, one should 
also explain the difference between the $S_{\phi K_S}$ and the other
$b\to s\bar ss$ processes, $S_{\eta^\prime K_S}$ and $-S_{K^+K^-
  K_S}$.  (Of course, the large statistical errors make any
conclusion premature).

\subsection{$B\to\psi\pi^0$ and $B\to D^{*+}D^{*-}$}
The $B\to\psi\pi^0$ and $B\to D^{*+}D^{*-}$ decays, related to the
$\bar b\to\bar cc\bar d$ quark transition, get contributions from both
tree and penguin diagrams, where the CKM combinations are of similar
magnitude, ${\cal O}(\lambda^3)$, but carry different
phases. Consequently, the deviation of $|\bar A/A|$ from unity can be
large. Since both $C$ and, in particular, $S$ can be large, CP
violation in mixing can be safely neglected.

\begin{figure}[htb]
  \includegraphics*[angle=90,width=\columnwidth]{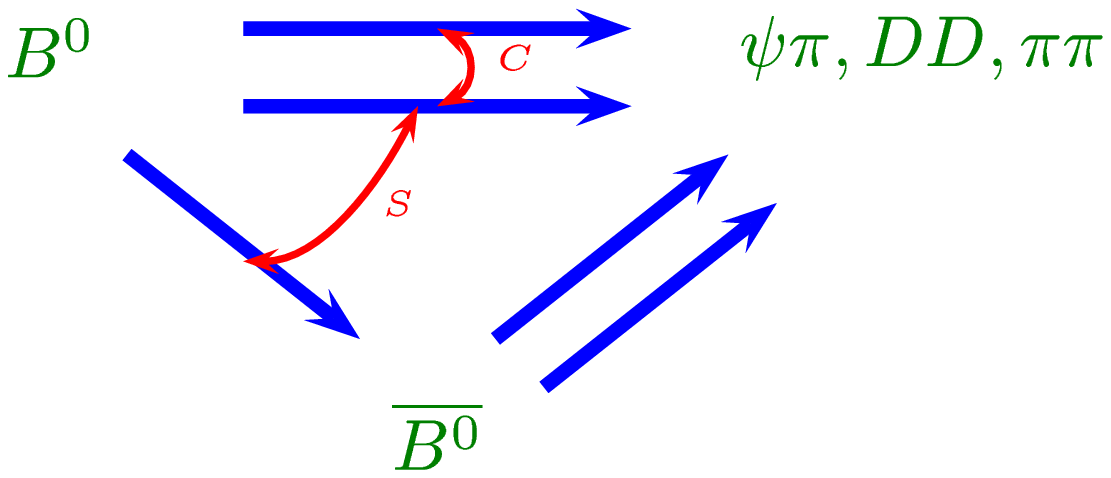}
\label{fig:btopipi}
\end{figure}

Averaging over the new Belle \cite{yamauchi} and Babar
\cite{karyotakis} results for the $\psi\pi^0$ mode, we obtain the
present world average, 
\begin{eqnarray}\label{psipiexp}
  C_{\psi\pi}&=&+0.31\pm0.29,\nonumber\\
  S_{\psi\pi}&=&-0.46\pm0.49.
\end{eqnarray}
The $D^{*+}D^{*-}$ state is not a CP eigenstate. The Babar collaboration
has, however, performed an angular analysis which separates the
CP-even and CP-odd components, with the following result for the
CP-even final state \cite{karyotakis}:
\begin{eqnarray}\label{ddexp}
  |\lambda_{(D^{*+}D^{*-})_+}|&=&0.98\pm0.27,\nonumber\\
  {\cal I}m\lambda_{(D^{*+}D^{*-})_+}&=&0.31\pm0.46.
\end{eqnarray}
Thus, CP violation in $b\to c\bar cd$ transitions has not yet been
observed.

The CKM dependence of the tree contribution to $\lambda$ in $b\to
  c\bar cd$ transitions is the same as that of $\lambda_{\psi K}$ of
eq. (\ref{lampks}). Loop contributions, however, modify the CKM
dependence. Defining $T$ and $P$ through
\begin{equation}\label{ptccd}
  A_f \equiv T_f V_{cb}^*V_{cd}+P_f V_{tb}^*V_{td},
\end{equation}
we obtain
\begin{equation}\label{lamccd}
  \lambda_{f_{\pm}(\bar cc\bar dd)}=\pm
  e^{-2i\beta}\left({1+(P_fR_t/T_f)e^{+i\beta}\over
      1+(P_fR_t/T_f)e^{-i\beta}}\right).
\end{equation}
Consequently, a violation of either $-S_{\psi\pi}=S_{\psi K_S}$ or
$-S_{(DD)_+}=S_{\psi K_S}$ will signal direct CP violation and, in
particular, a significant penguin contribution to the decay. (If the
violation is very strong, it might signal new physics. This statement
is particularly valid for the $DD$ mode, where the penguin
contribution is expected to be small.)

The measurements of the two modes suffer from large statistical
errors. At present, there is no evidence for either  $-S_{\psi\pi}\neq
S_{\psi K_S}$ or $-S_{D^*D^*}\neq S_{\psi K_S}$. We conclude that, at
present, there is no evidence for `penguin pollution' in these measurements.

\subsection{$B\to\pi\pi$}
The $B\to\pi\pi$ decay, related to the $\bar b\to\bar uu\bar d$ quark
transition, gets contributions from both tree and penguin diagrams,
where the CKM combinations are of similar magnitude, ${\cal
  O}(\lambda^3)$, but carry different phases. Consequently, the
deviation of $|\bar A/A|$ from unity can be large. Since both $C$ and,
in particular, $S$ can be large, CP violation in mixing can be safely
neglected.

The results of Belle (based on 41.8 fb$^{-1}$ \cite{Abe:2002qq} and
not updated in this conference) and Babar \cite{karyotakis} for CP
violation in $B\to\pi\pi$ suffer from large statistical errors and
are inconsistent with each other. It is, therefore, more prudent at
present to quote the separate results rather than average over them:
\begin{eqnarray}\label{pipiex}
  C_{\pi\pi}&=&\cases{-0.94^{+0.31}_{-0.25}\pm0.09&Belle,\cr
    -0.30\pm0.25\pm0.04&Babar,\cr}\nonumber\\
   S_{\pi\pi}&=&\cases{-1.21^{+0.38+0.16}_{-0.27-0.13}&Belle,\cr 
     +0.02\pm0.34\pm0.05&Babar.\cr}
 \end{eqnarray}

\begin{table*}[htb]
\caption{CP asymmetries in $B\to f_{\rm CP}$.}
\label{table:1}
\newcommand{\m}{\hphantom{$-$}}
\newcommand{\cc}[1]{\multicolumn{1}{c}{#1}}
\renewcommand{\arraystretch}{1.2} 
\begin{tabular}{@{}llllll}
\hline
$f_{\rm CP}$  & $b\to q\bar qq^\prime$ & SM & $S$ &
$C$ & $-\eta_{\rm CP}S\neq S_{\psi K}$?$^{(1)}$ \\
\hline
$\psi K_S$    & $b\to c\bar cs$ & $\sin2\beta$ & $+0.734\pm0.054$ &
$|\lambda|=0.95\pm0.04$ &  \\
$\phi K_S$    & $b\to s\bar ss$ & $\sin2\beta$ & $-0.39\pm0.41$ &
$+0.56\pm0.43$ & $2.7\sigma$  \\
$\eta^\prime K_S$ & $b\to s\bar ss$ & $\sin2\beta$ & $+0.76\pm0.36$ &
$-0.26\pm0.22$ & $-$  \\
$K^+K^-K_S{}^{(2)}$ & $b\to s\bar ss$ & $\sin2\beta$ & $-0.52\pm0.47$ &
$+0.42\pm0.37$ & $-$  \\
$\psi\pi^0$ & $b\to c\bar cd$ & $\sin2\beta_{\rm eff}$ &
$-0.46\pm0.49$ & $+0.31\pm0.29$ & $-$  \\
$D^{*+}D^{*-}{}^{(3)}$ & $b\to c\bar cd$ & $\sin2\beta_{\rm eff}$ & ${\cal
  I}m\lambda=+0.31\pm0.46$ & $|\lambda|=0.98\pm0.27$ &
$2.7\sigma$  \\
$\pi^+\pi^-$ & $b\to u\bar ud$ & $\sin2\alpha_{\rm eff}$ &
$-0.48\pm0.60$ & $-0.54\pm0.31$ & $-$  \\
\hline
\end{tabular}\\[2pt]
$^{(1)}\eta_{\rm CP}=+(-)1$ for CP even (odd) states.\\
$^{(2)}$Isospin analysis was used to argue that
$K^+K^-K_S$ is dominated by CP-even states.\\
$^{(3)}$Angular analysis was used to separate CP-even and CP-odd
$D^{*+}D^{*-}$ states.  
\end{table*}
 
Defining $T$ and $P$ through
\begin{equation}\label{ptuud}
  A_{\pi\pi} \equiv T_{\pi\pi} V_{ub}^*V_{ud}+P_{\pi\pi} V_{cb}^*V_{cd},
\end{equation}
we obtain
\begin{equation}\label{lampipi}
  \lambda_{\pi\pi}=
  e^{2i\alpha}\left({1+(P_{\pi\pi}/T_{\pi\pi}R_u)e^{+i\gamma}\over
      1+(P_{\pi\pi}/T_{\pi\pi}R_u)e^{-i\gamma}}\right).
\end{equation}
The (expected) violation of  $-S_{\pi\pi}=S_{\psi K_S}$ or of
$C_{\pi\pi}=0$ will signal direct CP violation. At present, however,
there is no evidence for either of these options. (The average of the
two measurements, with errors reflecting the inconsistency, is
$C_{\pi\pi}=-0.54\pm0.31$ and $S_{\pi\pi}=-0.48\pm0.60$.)

The CP asymmetries in $B\to\pi\pi$ decays are one of the most
interesting measurements anticipated in the $B$-factories because it
can potentially determine the angle $\alpha$, thus providing
yet another independent constraint on the unitarity triangle. As can
be seen from eq. (\ref{lampipi}), if the penguin contributions were
negligible, one would simply have ${\cal
  I}m\lambda_{\pi\pi}=\sin2\alpha$. The ratio $P_{\pi\pi}/T_{\pi\pi}$
is, however, non-negligible. To proceed, one can choose one of the
following options:

1. {\it A model independent determination of $\alpha$ from
  isospin analysis} \cite{Gronau:1990ka}. This method requires 
  measurements of various branching fractions and CP asymmetries and
  does not yet yield useful constraints.

2. {\it A model independent upper bound on the deviation of
      $\sin2\alpha_{\rm eff}\equiv{{\cal
          I}m\lambda_{\pi\pi}\over|\lambda_{\pi\pi}|}$ from
      $\sin2\alpha$} \cite{Grossman:1997jr,Charles:1998qx}:
\begin{equation}\label{groqui}
\sin^2(\alpha_{\rm eff}-\alpha)\leq{{\cal B}(B\to\pi^0\pi^0)\over
  {\cal B}(B^+\to\pi^+\pi^0)}\leq0.6.
\end{equation}
Thus $(\alpha_{\rm eff}-\alpha)$ in the range $50^o-130^o$ is
excluded, a useful though not very strong bound.

3. {\it A model dependent determination of $\alpha$ with a theoretical
  value for $P_{\pi\pi}/T_{\pi\pi}$}
  \cite{Beneke:2000ry,Gronau:2002qj}. In this 
  context we would like to mention that a large (small) value of
  $|C_{\pi\pi}|$ would (might) give evidence for a large (small)
  strong phase, in contrast to the prediction of
  \cite{Beneke:2000ry} (\cite{Keum:2000ph}).

In the future, measurements of CP violation in $B\to\rho\pi$ decays
will contribute to a model independent determination of $\alpha$
\cite{Lipkin:1991st,Snyder:1993mx,Quinn:2000by}. The first
experimental steps towards this goal have been reported in this
conference \cite{karyotakis}.

\subsection{Summary}

The results of the searches for CP asymmetries in $B$ decays into
final CP eigenstates are summarized in Table \ref{table:1}.
We can describe the emerging picture as follows:
\begin{itemize}
  \item CP violation has not yet been observed in $B$ decays other than 
      $B\to \psi K$. (The largest effect is at the $2.1\sigma$ level
      in $S_{\eta^\prime K_S}$.) 
  \item Direct CP violation has not yet been observed in $B$
    decays. (The largest effects are at the $2.7\sigma$ level in
    $S_{\phi K}-S_{\psi K}$ and in $S_{DD}+S_{\psi K}$.) 
  \item There is no evidence of new physics. (The largest effect that
    is inconsistent with the Standard Model prediction in a
    $2.7\sigma$ violation of $S_{\phi K}=S_{\psi K}$.)
  \item The measurements of branching ratios and CP asymmetries in
    $B\to\pi\pi$ decays are at a stage where restrictions on the
    CKM parameters and on hadronic parameters begin to emerge. The
    model independent constraints are still mild.
\end{itemize}

\section{New Physics Lessons}
CP violation is an excellent probe of new physics. First, the
uniqueness of the Kobayashi-Maskawa mechanism (with its single source
of CP violation) means that the mechanism is predictive and testable,
and that new sources of flavor and CP violation can induce
large deviations from the Standard Model predictions. Second,
the cleanliness of the theoretical interpretation of various CP
asymmetries means that signals of new physics will not be obscured 
by hadronic uncertainties.

The supersymmetric Standard Model provides an impressive example of
the rich possibilities for the physics of CP violation. The model
depends on 124 independent parameters, many of which are flavor
violating. In particular, 44 of the parameters are CP
violating. Consequently, Standard Model correlations
between CP violating observables may be violated, and Standard Model
zeros in CP asymmetries may be lifted. We use supersymmetry here as an
example of new physics that allows us to ask two questions:
\begin{itemize}
  \item What are the implications of existing measurements of CP
    violation on supersymmetric model building?
    \item What are the prospects that future measurements of CP
      violation will discover deviations from the Standard Model
      predictions?
    \end{itemize}

\subsection{SUSY model building}
If all flavor violating and CP violating parameters of the
supersymmetric Standard Model were of order one, then constraints from
flavor changing neutral current processes and from CP violation would
be violated by many orders of magnitude. This statement is best
demonstrated when comparing the supersymmetric contribution to the
imaginary part of the $K^0-\overline{K^0}$ mixing amplitude to the
experimental constraint (derived from $\Delta m_K\times\varepsilon$,
see for example \cite{Gabrielli:1995bd}):
\begin{eqnarray}\label{susyeps}
  {({\cal I}m M_{12}^K)^{\rm SUSY}\over({\cal I}m M_{12}^K)^{\rm
  EXPT}}&\sim& 10^8\ {m_Z^2\over\tilde m^2}\ {\Delta
  m^2_{\tilde Q}\over m^2_{\tilde Q}}\ {\Delta
  m^2_{\tilde D}\over m^2_{\tilde D}}\nonumber\\
&\times&{\cal I}m[(K^d_{LL})_{12}(K^d_{RR})_{12}],  
\end{eqnarray}
where $\tilde m$ is the scale of the soft SUSY breaking terms, $\Delta
m^2_{\tilde Q(\tilde D)}$ is the mass-squared difference between the
first two doublet (singlet) down squark generations, $m_{\tilde
  Q(\tilde D)}$ is their average mass, and $K^d_{LL(RR)}$ is the mixing
matrix for doublet (singlet) down squarks. Even when one allows for
squark and gluino masses close to $300\ GeV$ and for approximate
degeneracy induced by RGE effects ($\Delta m^2/m^2\sim0.15$), the
bound is violated by $4-5$ orders of magnitude. Consequently, $K$
physics has had a huge impact on SUSY model
building. Eq. (\ref{susyeps}) shows the various ways in which {\it the
  supersymmetric $\varepsilon$ problem} can be solved:
\begin{itemize}
\item Heavy squarks: $\tilde m\gg100\ GeV$;
\item Universality: $\Delta m^2_{\tilde s\tilde d}\ll\tilde m^2$;
\item Alignment: $|K^d_{12}|\ll1$;
\item Approximate CP: $\sin\phi\ll1$, where $\phi$ stands for a CP
  violating phase.
\end{itemize}
Any viable model of supersymmetry employs one of these options or some
combination of them.

The scenario where all CP violating phases are small has been well
motivated within the supersymmetric framework. The reason is that even
if one employs, say, exact universality to solve all supersymmetric
flavor-changing CP problems, there remains a supersymmetric
flavor-diagonal CP problem \cite{Dugan:1984qf}. This statement is
best demonstrated by comparing the supersymmetric contribution to the
electric dipole moment (EDM) of the neutron with the experimental bound:
\begin{equation}\label{susydn}
  {d_N^{\rm SUSY}\over6.3\times10^{-26}\ e\ {\rm cm}}\sim300\ 
  {m_Z^2\over\tilde m^2}\ \sin\phi_{A,B}.
\end{equation}
Here $\phi_A$ (related to the trilinear scalar couplings and gaugino
masses) and $\phi_B$ (related to the bilinear Higgs coupling, bilinear
Higgsino coupling and gaugino masses) are the two physical
flavor-diagonal phases that remain even in models of exact universality.
In this context, the measurement of the CP asymmetry in $B\to\psi
K_S$ decays has an immediate impact. Since the measured asymmetry is of order
one, it excludes the idea of approximate CP. To solve the
{\it supersymmetric EDM problem} one has to either assume that the squarks
(of at least the first two generations) are heavy or invoke special
mechanisms that suppress $\phi_A$ and $\phi_B$ but not all
flavor-changing CP phases. (It is still possible, but highly unlikely,
that the large $S_{\psi K}$ is induced by a small phase, if there are
fine-tuned cancellations between various contributions to the real
part of the $B^0-\overline{B^0}$ mixing amplitude.)

Since the measured value of $S_{\psi K}$ is consistent with the
Standard Model prediction, it is also consistent with models of exact
universality. Such models provide a well motivated example of a class
of models where the CKM matrix is the only source of
flavor changing and CP violating couplings. This framework is often
called `{\it minimal flavor violation}' (MFV)
\cite{Ciuchini:1998xy,Ali:1999we,Buras:2000dm} (for recent analyses,
see \cite{D'Ambrosio:2002ex,Buras:2002yj}). In contrast, 
the consistency of the measured $S_{\psi K}$ with the Standard Model
prediction provides interesting constraints on models with genuinely
supersymmetric sources of flavor and CP violation
\cite{Becirevic:2001jj}. In particular, models of heavy squarks, where
flavor violation is only very mildly suppressed \cite{Cohen:1996sq},
are disfavored.

The search for mixing and CP violation in the neutral $D$ system is
the most promising probe of supersymmetric models with
alignment. Here, there are two important points concerning CP
violation. First, given that it is not impossible for the Standard
Model contributions to induce mixing close to present bounds
\cite{Falk:2001hx}, CP violation in mixing is crucial to make a
convincing case for new physics \cite{Bergmann:2000id}. Second, the
possible presence of CP violation should be taken into account when
translating the experimental bounds into constraints on new physics. 
Specifically, one obtains \cite{Raz:2002ms}
\begin{equation}\label{motdcpv}
  |M_{12}^D|\lsim5.4\times10^{-11}\ MeV,
\end{equation}
compared to the PDG bound of $2.3\times10^{-11}\ MeV$ which assumes
vanishing weak and strong phases. The bound of eq. (\ref{motdcpv})
implies a severe constraint on squark masses in alignment models
\cite{Nir:2002ah}.

The mechanism by which supersymmetric flavor and CP violation are
suppressed relates directly to the mechanism of dynamical supersymmetry
breaking. In this sense, if supersymmetry is discovered, measurements
of CP violating observables will play a crucial role in understanding
the full high-energy theory \cite{Dine:2001ne,Masiero:xj}. Future
searches for EDMs are important in this context. While supersymmetric
models with only the Kobayashi-Maskawa phase as a source of CP
violation (such as certain models of gauge mediation) typically give
$d_N\lsim10^{-31} e$ cm, those where supersymmetry breaking is
communicated at or close to the Planck scale typically give
$d_N\gsim10^{-28} e$ cm.   
  
\subsection{SUSY CP violation in $B\to\phi K_S$?}
We choose the question of whether supersymmetry can significantly
modify the CP asymmetry in the $B\to\phi K_S$ decay to demonstrate
several points that have more general applicability.

Supersymmetry contributes to the $B\to\phi K_S$ decay via, for
example, gluonic penguin diagrams with intermediate squarks and
gluinos \cite{Bertolini:1987pk,Lunghi:2001af}.

\begin{figure}[htb]
  \includegraphics*[angle=90,width=\columnwidth]{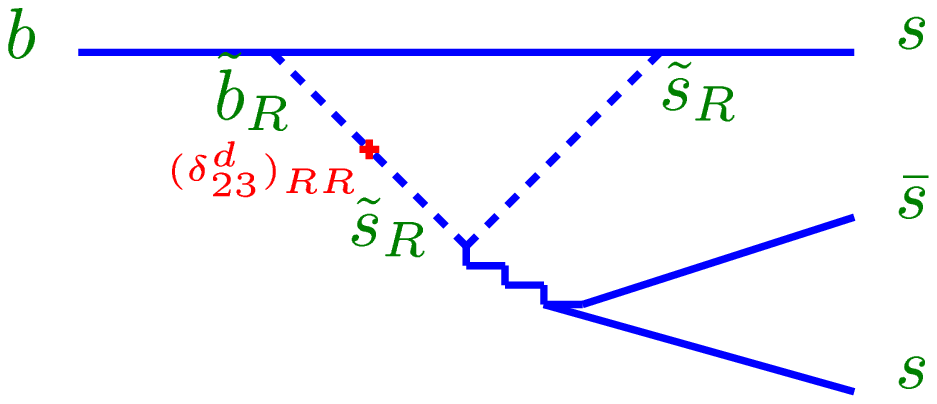}
\label{fig:btophik}
\end{figure}

Two crucial questions, regarding the size of this effect, come
immediately to mind:

$\bullet$ Is it still possible that new physics has an ${\cal O}(1)$
effect on $S_{\phi K}$ when the measured value of $S_{\psi K}$ is so
close to the Standard Model prediction?

The answer to this question is positive. Assuming that the consistency
is not accidental, it teaches us that the effect of new, CP violating
physics in $b\to d$ transitions is small. The $B\to\phi K_S$ decay
involves, however, not only $B^0-\overline{B^0}$ mixing but also the
$b\to s\bar ss$ decay. There is no similar indication for the
smallness of new, CP violating effects in $b\to s$ transitions.

In the language of supersymmetry, the $S_{\psi K}$ constraint applies
to $\delta^d_{13}$ \cite{Becirevic:2001jj} while $B\to\phi K_S$
depends also on $\delta^d_{23}$.

$\bullet$ Is it still possible that new physics has an ${\cal O}(1)$
effect on $S_{\phi K}$ when the measured value of ${\cal B}(B\to
X_s\gamma)$  is close to the Standard Model prediction?

The answer to this question is, again, positive. The $b\to s\gamma$
rate teaches us that the effects of new, helicity changing
physics in $b\to s$ transitions is small. The $B\to\phi K_S$ decay
involves, however, also helicity conserving contributions, which are
much more weakly constrained \cite{Borzumati:1999qt,Besmer:2001cj}.

In the language of supersymmetry, the ${\cal B}(B\to X_s\gamma)$
constraint is significant for $\delta^d_{LR}$ while $B\to\phi K_S$
depends also on $\delta^d_{LL}$ and $\delta^d_{RR}$.

Having answered these two generic questions in the affirmative, one
may ask a third question that is more specific to supersymmetry:

$\bullet$ Are there well motivated supersymmetric models where new, CP
violating effects in $b\to s$ transitions could be particularly large?

We are aware of, at least, two classes of models where this is indeed
the case. First,
in models where an Abelian horizontal symmetry determines the flavor
structure of both the quark Yukawa matrices and the squark
mass-squared matrices, the supersymmetric mixing angles are related to
the quark parameters
\cite{Leurer:1993gy,Arhrib:2001jg,Chua:2001dd}. In particular, we have 
\begin{equation}\label{abebs}
  (\delta^d_{RR})_{23}\sim{m_s/m_b\over |V_{cb}|}={\cal O}(1).
\end{equation}
Note, however, that if alignment is to solve the
supersymmetric flavor problem without any squark degeneracy,
$(\delta^d_{RR})_{23}$ must be suppressed compared to the estimate
(\ref{abebs}). Similarly, in models where a
non-Abelian horizontal symmetry determines all flavor parameters,
$(\delta^d_{RR})_{23}\ll1$ and so is the difference between $S_{\phi
  K}$ and $S_{\psi K}$ \cite{Barbieri:1997tu,Barbieri:1997kq}.

Second, in supersymmetric GUT theories, the $b_R-s_R$ mixing is
related to the 
$\nu_\mu-\nu_\tau$ mixing. The latter is required to be of order one
to explain the atmospheric neutrino measurements. This intriguing
relation has particularly interesting consequences in the framework of 
$SO(10)\to SU(5)$ theories, where one obtains
\cite{Moroi:2000tk,Chang:2002mq}  
\begin{equation}\label{gutbs}
  (\delta^d_{RR})_{23}\sim {15\over8\pi^2}\log{M_{\rm Pl}\over
  M_{10}}={\cal O}(0.5). 
\end{equation}

\subsection{Discussion}
\begin{figure*}[htb]
  \includegraphics*[width=0.33\textwidth]{rhoeta_K.eps}\hfill
  \includegraphics*[width=0.33\textwidth]{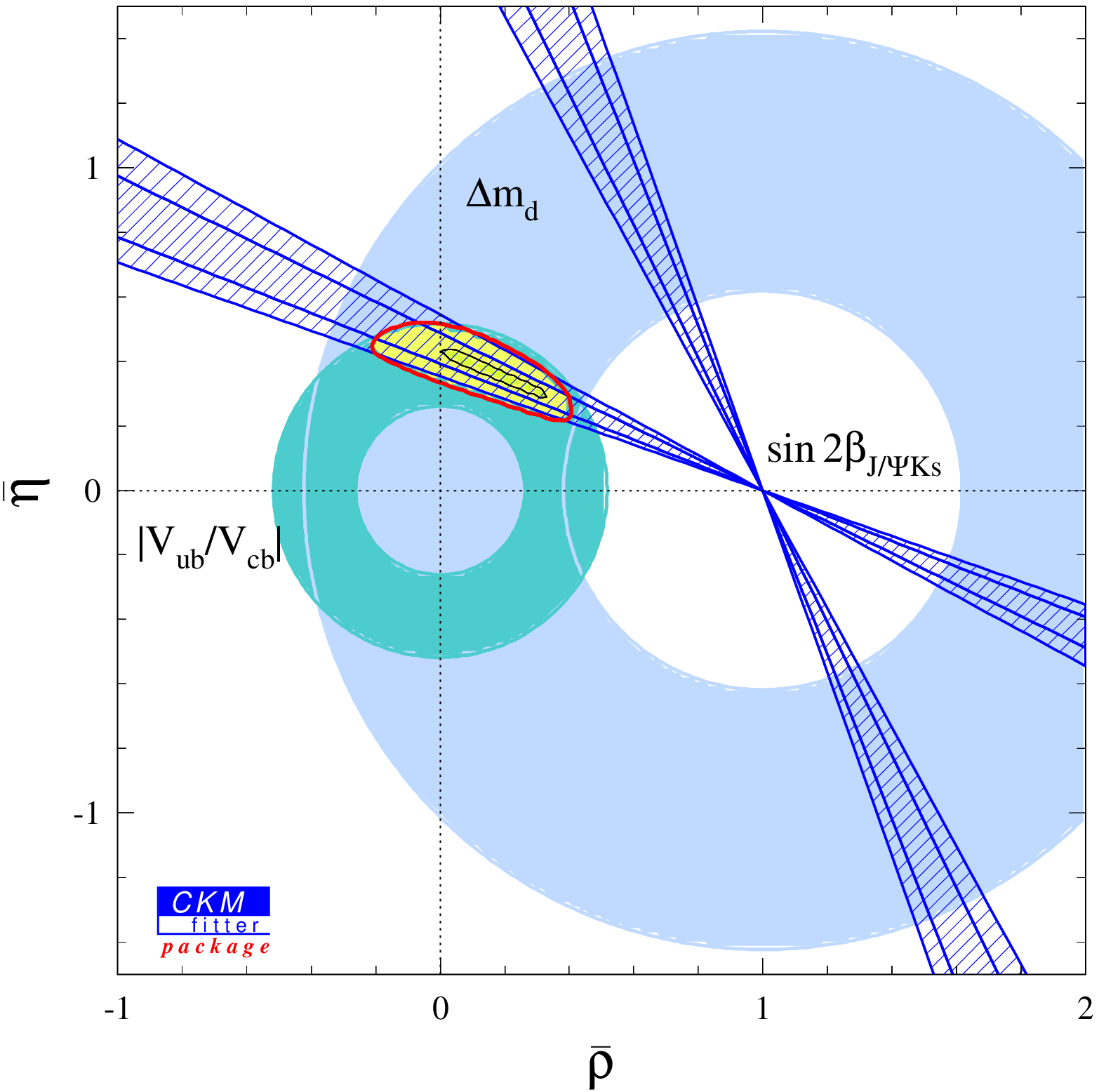}\hfill
  \includegraphics*[width=0.33\textwidth]{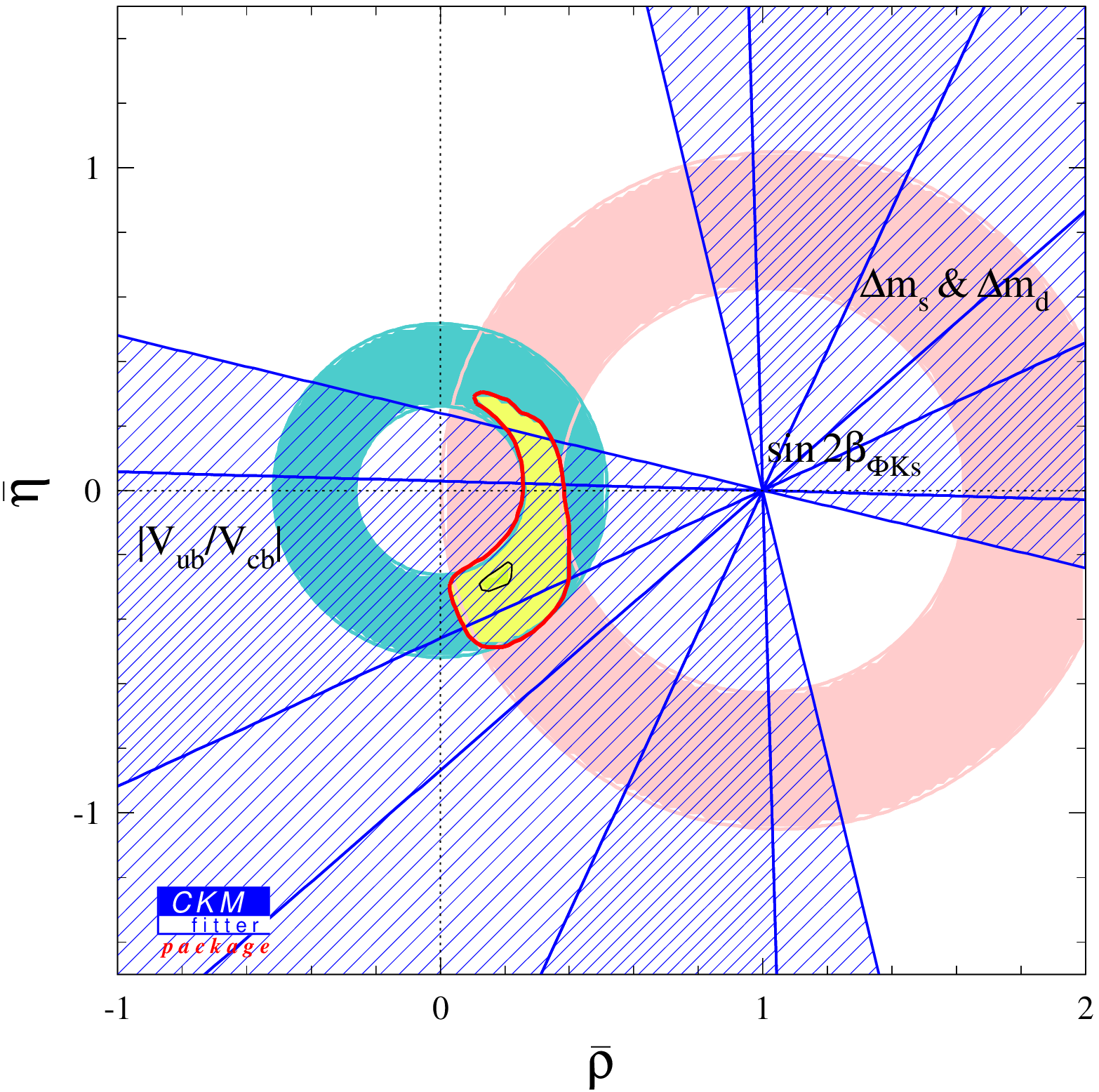}
\caption{Unitarity triangle constraints from tree level decays and
  from (left) $s\to d$, (center) $b\to d$, and (right) $b\to s$
  loop processes.} 
\label{fig:sdbdbs}
\end{figure*}

The general conclusion of the discussion of supersymmetry is that, if
the consistency between experiment and the Standard Model concerning
$S_{\psi K}$ is not accidental, then large ({\it i.e.} $\gg20\%$) CP
violating effects in $b\to d$ transitions are disfavored. In contrast,
the constraints on new CP violating effects in $b\to s$ transitions
are much milder.

We would like to demonstrate this point by performing a premature but,
hopefully, thought-provoking exercise. Instead of the ``$B$-triangle''
and ``$K$-triangle'' of Figure \ref{fig:bkut}, we
now divide the loop processes (sensitive to new physics) into three
classes. First, those involving $s\to d$ flavor-changing neutral
current (FCNC) transitions. This triangle is identical to the
$K$-triangle presented in Figure \ref{fig:bkut}. Second, processes
involving $b\to d$ FCNC transitions. Here we include only $\Delta m_B$
and $S_{\psi K}$. Finally, the constraints that arise from the lower
bound on $\Delta m_{B_s}/\Delta m_{B_d}$ and from the
bounds on $S_{\phi K_S}$. Both constraints involve, in addition to the
$B^0-\bar B^0$ mixing amplitude, $b\to s$ FCNC transitions. The
results of this exercise are presented in Figure \ref{fig:sdbdbs}.

There is one solid statement that can be made on the basis of this
presentation: {\it There is still a lot to be learnt from future
measurements}. In particular, the probing of new physics in FCNC and in
CP violating processes will become much more sensitive when the
following is achieved:

$\bullet$ The $b\to d$ triangle: The experimental accuracy and
theoretical understanding of $B$ decays that depend on $\alpha$,
$\gamma$, $|V_{ub}|$ and $|V_{cb}|$ improve.

$\bullet$ The $s\to d$ triangle: ${\cal B}(K^+\to\pi^+\nu\bar\nu)$ is
measured more precisely and the CP violating ${\cal
  B}(K_L\to\pi^0\nu\bar\nu)$ is measured
\cite{Buchalla:1993bv,Buras:1994rj,Grossman:1997sk,D'Ambrosio:2001zh}.

$\bullet$ The $b\to s$ triangle: The CP asymmetries in $B\to\phi K_S$
and other $b\to s\bar ss$ processes are measured more precisely and
$\Delta m_{B_s}$ and CP asymmetries in $B_s$ decays are measured.

$\bullet$ Standard Model zeros: The experimental sensitivities to CP
violation in $D^0-\overline{D^0}$ mixing and to EDMs improve. 

\section{Conclusions}

We have made a significant progress in our understanding of CP
violation. For the first time, we are able to make the following
statement based on experimental evidence:

$\bullet$ {\it Very likely, the Kobayashi-Maskawa mechanism is the dominant
  source of CP violation in flavor changing processes.}

One consequence of this development is the following:

$\bullet$ We are leaving the era of hoping for new physics {\it
  alternatives} to the CKM picture of flavor and CP violation.

In particular, the superweak scenario is excluded by the observation
 of direct CP violation in $K\to\pi\pi$ decays, and the scenario of
 approximate CP is excluded by the observation of an order
 one CP asymmetry in $B\to\psi K_S$ decays.

$\bullet$ We are entering the era of seeking for new physics {\it
  corrections} to the CKM picture.

This effort would require a broad experimental program, and
improvements in both the experimental accuracy and the theoretical
cleanliness of CKM tests.

One has also to bear in mind that rather large corrections are still
possible in $\Delta m_{B_s}$, in CP asymmetries in $B_s$ decays, and
in CP asymmetries in $B$ decays related to $\bar b\to \bar ss\bar s$
transitions. 

\bigskip
\leftline{\bf Acknowledgments}

I am grateful to Sandrine Laplace for working out the CKM fits,
for producing the figures of the unitarity triangle constraints, and
for many useful discussions. I thank Adam Falk, Yuval Grossman, Zoltan
Ligeti and Guy Raz for their help and advice in preparing this talk. 



\end{document}